\documentclass[manuscript,screen]{acmart}
\AtBeginDocument{%
  \providecommand\BibTeX{{%
    \normalfont B\kern-0.5em{\scshape i\kern-0.25em b}\kern-0.8em\TeX}}}

\setcopyright{acmcopyright}
\copyrightyear{2023}
\acmYear{2023}
\acmDOI{XXXXXXX.XXXXXXX}

\acmConference[CHI '23]{CHI 2023 Workshop on Building Credibility, Trust, and Safety on Video-Sharing Platforms, Hamburg, Germany}{April 23--28, 2023}{New Orleans, LA, USA}
\acmBooktitle{CHI 2023 Workshop on Building Credibility, Trust, and Safety on Video-Sharing Platforms, April 23--28, 2023, Hamburg, Germany}

\begin{document}

\title[The Challenges of Studying Misinformation on Video-Sharing Platforms]{The Challenges of Studying Misinformation on Video-Sharing Platforms During Crises and Mass-Convergence Events}

\author{Sukrit Venkatagiri}
\authornote{All authors contributed equally to this research.}
\email{sukritv@uw.edu}

\author{Joseph S. Schafer}
\authornotemark[1]
\email{schaferj@uw.edu}
\affiliation{%
  \institution{Center for an Informed Public, Human Centered Design and Engineering, University of Washington}
  \city{Seattle}
  \state{Washington}
  \country{USA}
}

\author{Stephen Prochaska}
\authornotemark[1]
\email{sprochas@uw.edu}
\affiliation{%
  \institution{Center for an Informed Public, The Information School, University of Washington}
  \city{Seattle}
  \state{Washington}
  \country{USA}
}

\renewcommand{\shortauthors}{Venkatagiri et al.}

\begin{abstract}
Mis- and disinformation can spread rapidly on video-sharing platforms (VSPs). Despite the growing use of VSPs, there has not been a proportional increase in our ability to understand this medium and the messages conveyed through it. In this work, we draw on our prior experiences to outline three core challenges faced in studying VSPs in high-stakes and fast-paced settings: (1) navigating the unique affordances of VSPs, (2) understanding VSP content and determining its authenticity, and (3) novel user behaviors on VSPs for spreading misinformation. By highlighting these challenges, we hope that researchers can reflect on how to adapt existing research methods and tools to these new contexts, or develop entirely new ones.
\end{abstract}



\maketitle

\section{Introduction}
\begin{quote}
``\textit{The medium is the message. This is merely to say that the personal and social consequences of any medium --- that is, of any extension of ourselves --- result from the new scale that is introduced into our affairs by each extension of ourselves, or by any new technology.}''
— Marshall McLuhan in Understanding Media: The Extensions of Man \cite{mcluhan1994understanding}
\end{quote}

Video-sharing platforms (VSPs) enable individuals to better express themselves online but also enable the rapid spread of mis- and disinformation that can undermine trust in institutions, governments, and one another \cite{hussein2020measuring}. Despite the growing use of VSPs, there has not been a proportional increase in our ability to understand this medium and the messages conveyed through it \cite{niu2023building}. 

Building on our experiences of rapidly responding to misinformation during crises and mass-convergence events \cite{eip2021longfuse}, we outline three core challenges faced in studying VSPs in these high-stakes and fast-paced settings: (1) navigating the unique affordances of VSPs, (2) understanding VSP content and determining its authenticity, and (3) novel user behaviors on VSPs for spreading misinformation. By highlighting these challenges, we hope that researchers can reflect on how to adapt existing research methods and tools to these new contexts, or develop entirely new ones.

\section{Affordance Challenges}
Our first set of challenges related to studying misinformation on VSPs is related to the unique affordances that they offer. Specifically, the affordances of VSPs present five challenges for rapid response misinformation research: (1) the number of informational channels per post, (2) the increased importance of ephemeral video content, (3) the limited visibility of social interaction networks on VSPs, (4) the limited visibility into the user experience of recommendation feeds of content, and (5) the limited API support for researcher access to VSP data.

First, the affordances of video allow for multiple information channels where misinformation could be spreading. As acknowledged by \citet{niu2023building}, these locations include the audio of a video, text displayed in a video (either through platform text features or through other displayed text), alt-text and subtitles for a video, post captions, and comments. Some VSPs like TikTok and Instagram Reels even augment this further, through things like “audio” names (background sounds which can be used across multiple videos). In comparison to text-based platforms like Twitter, this is a dramatic increase in the number of informational channels per post in which misinformation might spread, and the number of forms of data to which rapid-response misinformation researchers need to pay attention. 

The ephemeral nature of content on many VSPs also complicates rapid-response misinformation research, at both the methodological and ethical levels. Methodologically, accessing data in live streamed or temporarily available content is difficult, and in the live stream case may be impossible to retrieve after the fact, depending on the platform. Ethically, data which is only temporarily “public” adds further complications to what researchers ought to examine when working in the public interest, a question which has been explored by researchers before \citep[e.g.,][]{banchik2021disappearing, bipat2017live, boccia2021below}. 

Another challenge of conducting rapid-response misinformation research is that networks of linked content are much more difficult to trace. While this existed with multimedia elements on other platforms, these problems are exacerbated on VSPs. For example, TikTok and Instagram reels have platform features (called stitches and duets on TikTok) which are somewhat analogous to the ‘quote-tweet’ affordance on Twitter, an important method of information recontextualization and community interaction. While workarounds exist to get access to all the stitches and duets of a creator, there currently is no way to easily filter these interactions for specific videos. This is in stark contrast to Twitter, where this information is made easily available. Understanding how information flows and is retransmitted on video platforms is important to learning how misinformation might spread, but is currently not possible with existing platform features. 

However, even if these reposting behaviors were made easily accessible, understanding how users actually experience content and misinformation on VSPs like TikTok is far more difficult. While this is not unique to TikTok (particularly as other platforms add more algorithmic feeds as well, like Twitter’s For You feed or Instagram’s Reels feed), the primary mode of seeing content as a user does not come from who a user is following but rather from algorithmic recommendations. While we can somewhat approximate what kinds of content Twitter users might be exposed to from looking at their following networks, on TikTok, Instagram Reels, or other feed-based VSPs, this is not really possible to approximate without significant visibility into the workings of the recommendation algorithm. Adapting algorithmic auditing approaches e.g. \cite{hussein2020measuring} or bot account feed-simulating approaches e.g. \cite{bandy2021more} may be effective for understanding this broader landscape, but quickly identifying the communities a particular piece of misinformation content is spreading in becomes much more difficult with recommendation- and personalization-based content discovery, especially in rapid-response contexts.

One other affordance-related challenge of studying misinformation on video platforms in a rapid-response context is the lack of API support for researchers studying these topics. While some platforms have announced APIs \cite{tiktok2023supporting} or have limited existing APIs \cite{youtube2023API}, in general these tools are slim to non-existent. To do large-scale, data-driven research on these topics, as has been common in misinformation research thus far (e.g. \cite{kennedy2022repeat, vosoughi2018spread}), API systems for accessing VSP data are vital but currently not sufficient. 

\section{Content Understanding and Authenticity Challenges}
Apart from challenges posed due to the affordances of VSPs, there are four challenges posed in understanding the content itself: (1) limited capabilities of text and image-based search, (2) scaling up current analysis methods and moving beyond analyzing metadata, text, and transcripts, (3) determining veracity of claims made and authenticity of the video, and (4) determining a user’s intent in uploading a video and how it differs from viewers’ interpretations of that message.

First, beyond the lack of APIs — which makes it difficult for researchers to retrieve videos \cite{freelon2018computational} — it is difficult to identify relevant videos. Current search methods are largely based on video titles and metadata fields, and not the content itself — such as what is said within a video or what can be seen. For instance, it would be computationally expensive to conduct a “reverse video search” or to query a VSP for “all videos containing [an object]” \cite{apostolidis2019video}.

Second, even if all relevant videos could be obtained, researchers may face difficulties in analyzing them. The vast amounts of content uploaded online poses a challenge of scaling up current qualitative and quantitative analysis methods. For example, on YouTube alone, over 500 hours of video is uploaded every minute \footnote{https://blog.youtube/press/}. Qualitative methods, though beneficial in providing a deeper understanding of videos, are difficult to scale up to analyze hundreds of hours of video. Automated techniques may speed up analysis, but are largely limited to analyzing text of transcribed audio or are computationally expensive \cite{apostolidis2019video}. Many existing automated tools and techniques also require tuning of parameters or further training of models to be suitable for a particular research project. These temporal and computational limitations make it difficult to study video content during crises and other mass-convergence events that necessitate rapid analysis and response \cite{matatov2022stop}. 

A third challenge is determining the veracity of claims made within the video and the authenticity of the video itself. Determining the veracity of claims — or fact-checking videos — is a time-consuming process that requires making sense of disparate sources of information \cite{venkatagiri2019groundtruth}. Even if the veracity of claims made within a video, the authenticity of the video itself may come into question. The use of manipulated video, as well as synthetic or AI-generated video further add further complexity to the study of video content. Thus, not only would the claims themselves need to be verified, but the authenticity of individuals as well as other audio or visual content present within the video would need to be verified. The C2PA’s provenance pipeline is one potential solution \cite{rosenthol2022c2pa}, but not all video content would leverage C2PA’s approach.

Fourth, while fact-checking is largely an objective task, determining the intent behind a video is much more subjective \cite{zimmermann2020influencers}. For example, it may be difficult to determine why a user created, uploaded, or shared a video, without the user explicitly stating why. Potential reasons include making others aware of an event, criticizing or supporting a cause, or intentionally spreading misinformation, among others. While already difficult for text-based content, analyzing the intent behind video content may be even more difficult, because videos are multi-dimensional — allowing for greater variation in researchers’ and audiences’ interpretations \cite{zhou2022fake}. In this way, the message received by a user’s audience(s) may be different from a researcher’s interpretation of the intended message of a video (e.g., “dog whistling” or use of coded or suggestive language). VSP researchers should consider developing analytical frameworks to determine the intent behind a video being created or shared.

\section{Behavioral Dynamics Challenges}
The final group of challenges for researchers we have identified are those presented by the behaviors of the users of VSPs: (1) distinguishing between behaviors unique to VSPs and those that are common across platforms and mediums, (2) platform migration (where users move from one VSP to another), (3) accounting for audience folk theories of moderation and how those theories impact audience (and content creators’) behaviors, (4) obstacles presented by the strategic dissemination of inauthentic content, particularly synthetic media, and (5) take into account the impacts of proposed moderation strategies on communities that are disproportionately impacted by algorithmic moderation.

The first challenge in analyzing VSPs is that they have unique affordances, such as the ability to perform “duets” on TikTok, that enable unique behaviors such as amplifying content from a more extreme user while maintaining distance from the content itself. Disentangling novel VSP behaviors from their impacts will be a key area for future inquiry.

A second challenge that is common across mediums is the issue of platform migration. Many users will naturally utilize multiple platforms as they engage in online communities, making a full picture of their engagement difficult. This is exacerbated in communities that are the targets of perceived moderation efforts, such as conservative-leaning audiences discussing claims of election fraud. Members of these, and similar, communities often intentionally create content on low-moderation platforms like Rumble and then link to that content from platforms like Instagram, where the potential audience is much larger. There are also instances where users simply stop engaging on a platform where they perceive they are being moderated. This migration creates a major barrier for researchers who are seeking to understand how audiences engage with VSPs, especially when creators are already going to great lengths to avoid being seen by anyone other than their target audience.

The third challenge is the adversarial nature of some audiences’ behaviors as they attempt to avoid moderation. VSPs share many similar dynamics with more traditional social media platforms, where content creators and consumers often work together to develop “folk theories” about platform moderation and algorithmic amplification \cite{moran2022folk}. As their understanding of perceived algorithmic dynamics increases, they actively change their behavior to both ensure that content reaches a larger audience and to avoid moderation or perceived “shadowbanning.” In the realm of misinformation and particularly disinformation, this creates an adversarial environment where those interested in spreading their message are actively trying to take advantage of platform affordances to achieve their goals. As VSPs continue to gain popularity, content creators will likely continue to adapt their behaviors based on their shifting understandings of platform dynamics. This means that researchers need to grapple with not just the ephemerality of content, but also the ephemerality of community behaviors, which often changes at a pace faster than methodical research can keep up with. 

Fourth, even though the barrier to entry for coordinated inauthentic behavior and astroturfing campaigns is higher in VSPs, as AI-generated content becomes more sophisticated and more accessible to a larger range of users, experts across disciplines have warned (e.g. \cite{chesney2019deep}) that we will likely be seeing more of this type of behavior on VSPs, making it even more difficult for researchers to determine between authentic and inauthentic behaviors. At the time of writing, it is more difficult to scale inauthentic content on VSPs than on a text-based platform because of the relative complexity of video data over tweets, for example. However, as AI tools for the generation of deepfakes and other synthetic media become increasingly available, content creators interested in promoting their perspective by any means available are likely to start disseminating misleading deepfakes in larger quantities than have been seen thus far. We have recently seen the first use of a deepfake in an information operation \cite{graphika_2023}, and synthetic media for purposes other than mis- and disinformation have recently gained traction on VSPs \cite{cnn2021deepfake, insider2023voices}.

Finally, many of the strategies we have mentioned thus far are also utilized by diverse audiences and are not exclusively the domain of those engaged in the spread of disinformation or other problematic content. Members of historically marginalized communities have developed their own understandings of how algorithmic amplification and moderation works. As they are often the targets of algorithmic moderation even when their content does not violate platform policy, they have adjusted their behaviors in ways that may --- at a first glace --- appear to similar to coordinated and/or inauthentic behavior \cite{noble2018algorithms, bucher2012want, cotter2019playing}. This means that researchers need to develop a more nuanced understanding of (and methods of detecting)``problematic'' behaviors while avoiding further stifling the content and voices of historically marginalized content creators.

\section{Conclusion}
VSPs can be a vector for the rapid spread of mis- and disinformation during crises and mass-convergence events — necessitating a rapid response. Here, we preliminarily offer three sets of challenges to conducting rapid-response misinformation on VSPs, along three dimensions: (1) video-sharing platforms’ affordances, (2) content understanding and authenticity, and (3) behavioral dynamics of users. Revising old practices and developing new tools will be essential to addressing these challenges and promoting an informed public.

\begin{acks}
The authors are supported in part by the University of Washington Center for an Informed Public and the John S. and James L. Knight Foundation. Additional support was provided by the Election Trust Initiative and Craig Newmark Philanthropies. Joseph S. Schafer is also a recipient of an NSF Graduate Research Fellowship, grant DGE-2140004. Any opinions, findings, and conclusions or recommendations expressed in this material are those of the authors and do not necessarily reflect the views of the above supporting organizations or the National Science Foundation.
\end{acks}

\bibliographystyle{ACM-Reference-Format}
\bibliography{sample-base}

\end{document}